\documentstyle[sprocl,psfig]{article}

\newcommand {\ignore}[1]{}

\newcommand{\bc}{\begin{center}}
\newcommand{\ec}{\end{center}}

\def\ifmath#1{\relax\ifmmode #1\else $#1$\fi}

\def\3quarter{{\textstyle{3 \over 4}}}

\def\lf{\leaders\hbox to 1em{\hss.\hss}\hfill}
\def\e6{$E(6)$}
\def\10{$SO(10)$}
\def\21{$SU(2) \otimes U(1) $}

\def\422{$SU(4) \otimes SU(2) \otimes SU(2)$}
\def\321{$SU(3) \otimes SU(2) \otimes U(1)$}

\def\nt{\hbox{$\nu_\tau$ }}

\def\mnt{\hbox{$m_{\nu_\tau}$ }}

\def\eq#1{{eq. (\ref{#1})}}

\let\vev\VEV

\def\lsim{\raise0.3ex\hbox{$\;<$\kern-0.75em\raise-1.1ex\hbox{$\sim\;$}}}
\def\gsim{\raise0.3ex\hbox{$\;>$\kern-0.75em\raise-1.1ex\hbox{$\sim\;$}}}

\def\beq{\begin{equation}}
\def\eeq{\end{equation}}
\def\bef{\begin{figure}}
\def\eef{\end{figure}}
\def\bet{\begin{table}}
\def\eet{\end{table}}
\def\bea{\begin{eqnarray}}
\def\ba{\begin{array}}
\def\ea{\end{array}}
\def\bi{\begin{itemize}}
\def\ei{\end{itemize}}
\def\ben{\begin{enumerate}}
\def\een{\end{enumerate}}

\def\eea{\end{eqnarray}}
\def\n.c.#1#2#3{         {\it Nuovo Cim. }{\bf #1} (19#2) #3}
\def\r.n.c.#1#2#3{       {\it Riv. del Nuovo Cim. }{\bf #1} (19#2) #3}

\def\be{\begin{equation}}
\def\ee{\end{equation}}
\def\bea{\begin{eqnarray}}
\def\eea{\end{eqnarray}}

\begin{document}

\title{SUPER-GRAVITY UNIFICATION WITH BILINEAR R--PARITY VIOLATION}

\author{J. W. F. VALLE}

\address{ Departamento de F\'\i sica Te\'orica, IFIC-CSIC, 
Univ. de Valencia\\ 
Burjassot, Valencia 46100, Spain\\ 
http://neutrinos.uv.es} 

\maketitle\abstracts{Bilinear R--parity violation (BRpV) provides the
simplest and most meaningful way to include such effects into the
Minimal Supersymmetrical Standard Model (MSSM). It is defined by a
quadratic superpotential term $\epsilon L H$ which mixes lepton and
Higgs superfields and mimics the effects of models with spontaneous
breaking.  I review some of its main features and show how large
$\epsilon$ values can lead to a small neutrino mass radiatively,
without any fine-tuning. I discuss the effect of BRpV on gauge and
Yukawa unification, showing how bottom--tau unification can be
achieved at any value of $\tan\beta$. However, for very large m
$\epsilon$ values the large $\tan\beta$ solution is ruled out.  }

\section{Introduction}
 
Although the Standard Model (SM) works well in describing the
phenomenology of the strong and electroweak interactions of the known
particles it leaves unanswered some theoretical issues such as the
hierarchy problem and the unification of the gauge couplings at
$M_{GUT}$.  These provide strong impetus to the study of
supersymmetric extensions of the SM, the simplest being the Minimal
Supersymmetric Standard Model (MSSM) \cite{MSSM}.  In this case one
can show that the unification of gauge couplings at $M_{GUT}$
\cite{gaugeUnif,gaugUnifRecent} occurs for acceptable of the effective
Supersymmetry breaking as well as unification scales.

It is usual to assign to SM state an R--Parity defined by
$R_p=(-1)^{3B+L+2S}$, where $B$ is the baryon number, $L$ is the
lepton number and $S$ is the spin. In this way, quarks, leptons and
Higgs bosons are R--even, and the supersymmetric particles are R--odd.
If R--Parity is conserved, then supersymmetric particles are produced
in pairs in the laboratory. In addition, the lightest supersymmetric
particle (LSP, the lightest neutralino) is stable.

In contrast, if R--Parity is violated then supersymmetric particles
can be singly produced, and the LSP decays into standard quarks and
leptons. Moreover, the LSP need not be the lightest neutralino.

\section{R--Parity Violation}

Possible terms in the superpotential which violate R--Parity are
\begin{equation} 
W_{R_p\!\!\!\!\!\!/}=\lambda''_{ijk}\widehat U_i\widehat D_j\widehat D_k+
\varepsilon_{ab}\left[
 \lambda'_{ijk}\widehat L_i^a\widehat Q_j^b\widehat D_k
+\lambda_{ijk}\widehat L_i^a\widehat L_j^b\widehat R_k
+\epsilon_i\widehat L_i^a\widehat H_2^b\right]\,,
\label{WnotRp}
\end{equation}
Such terms may arise as residues of unification, e.g. as gravitational
effects \cite{BJV}.  The first three terms are Trilinear R--Parity
Violation (TRpV) terms. Each of the generation indices $i,j,k$ runs
from 1 to 3, thus implying a very large number of arbitrary
parameters. It is impossible to provide a systematic way to analyze
the implications of TRpV, the best one can do is to consider one or
two $\lambda$'s different from zero at a time. Some of these couplings
are strongly restricted by proton stability and/or primordial baryon
asymmetry survival.

The fourth term in eq.~(\ref{WnotRp}) corresponds to Bilinear
R--Parity Violation (BRpV) \cite{e3others,BRpVtalk}, and involves only
three extra parameters, one $\epsilon_i$ for each generation. The
$\epsilon_i$ terms also violate lepton number in the $i$th generation
respectively. Models where R--Parity is broken spontaneously
\cite{SRpSB} through vacuum expectation values (vev) of right handed
sneutrinos $\vev{\tilde\nu^c}=v_R\neq0$ generate BRpV (and not
TRpV)\footnote{ Of course, this is true in the original basis. If we
rotate the Higgs and Lepton superfields then TRpV terms are generated,
as explained later.  }.  Such spontaneous R-Parity Violation scenarios
are also interesting from the point of view of the electroweak phase
transition and baryogenesis \cite{ewbaryo} as well as
phenomenologically, due to the existence of massless pseudoscalar
majoron \cite{SRpSB} which brings in the possibility of invisibly
decaying Higgs boson  \cite{inv}.

The $\epsilon_i$ parameters are then equal to some Yukawa coupling
times $v_R$. This provides the main theoretical motivation for
introducing explicitly BRpV in the MSSM superpotential.  From a
practical point of view it provides the most predictive approach to
the violation of R--Parity, which renders possible the systematic
study of its phenomenological implications \cite{BRpVtalkphen}.  Here
I will mention the most important features of this model.

For simplicity we set from now on $\epsilon_1=\epsilon_2=0$, in this
way, only tau--lepton number is violated. In this case, considering
only the third generation, the MSSM--BRpV has the following
superpotential
\begin{equation} 
W=\varepsilon_{ab}\left[
 h_t\widehat Q_3^a\widehat U_3\widehat H_2^b
+h_b\widehat Q_3^b\widehat D_3\widehat H_1^a
+h_{\tau}\widehat L_3^b\widehat R_3\widehat H_1^a
-\mu\widehat H_1^a\widehat H_2^b
+\epsilon_3\widehat L_3^a\widehat H_2^b\right]\,,
\label{eq:Wsuppot}
\end{equation}
where the first four terms correspond to the MSSM. The last term violates
tau--lepton number as well as R--Parity. 

The presence of the $\epsilon$ term in the superpotential implies that
the tadpole equation for the tau sneutrino is non--trivial, i.e, the
vacuum expectation value $\vev{\tilde\nu_{\tau}}=v_3/\sqrt{2}$ is
non--zero. This in turn generates more R--parity and tau lepton number
violating terms which, as we will see later, induce a tau neutrino
mass.

It has often been claimed, by looking at the last two terms in the
superpotential, that the BRpV term be rotated away from the
superpotential by a suitable choice of basis \cite{HallSuzuki}.  If
this were true the $\epsilon$ term would be unphysical.  Indeed,
consider the following rotation of the superfields 
\begin{equation}
\widehat H_1'={{\mu\widehat H_1-\epsilon_3\widehat L_3}\over{
\sqrt{\mu^2+\epsilon_3^2}}}\,,\qquad
\widehat L_3'={{\epsilon_3\widehat H_1+\mu\widehat L_3}\over{
\sqrt{\mu^2+\epsilon_3^2}}}\,.
\label{eq:rotation}
\end{equation}
The superpotential in the new basis is
\footnote{For three generations there is also a $\widehat R \widehat
L' \widehat L'$ term.}
\begin{equation} 
W=h_t\widehat Q_3\widehat U_3\widehat H_2
+h_b{{\mu}\over{\mu'}}\widehat Q_3\widehat D_3\widehat H'_1
+h_{\tau}\widehat L'_3\widehat R_3\widehat H'_1
-\mu'\widehat H'_1\widehat H_2
+h_b{{\epsilon_3}\over{\mu'}}\widehat Q_3\widehat D_3\widehat L'_3
\,,\label{WsuppotP}
\end{equation}
where $\mu'^2=\mu^2+\epsilon_3^2$. The first four terms are MSSM-like
terms and the last term violates the R--Parity defined in the new
basis.  Notice that, although the $\epsilon$ term disappears from the
superpotential in the new basis, R--Parity is reintroduced in the form
of TRpV.  Moreover, supersymmetry must be broken and this is
parametrized by soft supersymmetry breaking terms. The soft terms
which play an important role in BRpV are the following
\begin{equation}
V_{soft}=m_{H_1}^2|H_1|^2+M_{L_3}^2|\widetilde L_3|^2
-\left[B\mu H_1H_2-B_2\epsilon_3\widetilde L_3H_2+h.c.\right]+...
\label{SoftUnrot}
\end{equation}
where $m_{H_1}^2$ and $M_{L_3}^2$ are the soft masses corresponding to
$H_1$ and $\widetilde L_3$ respectively, while $B$ and $B_2$ are the
bilinear soft mass parameters associated to the next-to-last and last
terms in the superpotential in eq.~(\ref{eq:Wsuppot}). It is clear,
for example, that Higgs vacuum expectation values
$\vev{H_i}=v_i/\sqrt{2}$ induce a non-trivial tadpole equation and a
non-zero vev for the sneutrino through the $B_2$ term in
eq.~(\ref{SoftUnrot}).

The soft terms in the rotated basis are given by
\begin{eqnarray}
V_{soft}&=&
{{m_{H_1}^2\mu^2+M_{L_3}^2\epsilon_3^2}\over{\mu'^2}}|H'_1|^2
+{{m_{H_1}^2\epsilon_3^2+M_{L_3}^2\mu^2}\over{\mu'^2}}|\widetilde L'_3|^2-
{{B\mu^2+B_2\epsilon_3^2}\over{\mu'}}H'_1H_2
\nonumber\\&&
+{{\epsilon_3\mu}\over{\mu'^2}}(m_{H_1}^2-M_{L_3}^2)\widetilde L'_3H'_1
+{{\epsilon_3\mu}\over{\mu'}}(B_2-B)\widetilde L'_3H_2+h.c.
+...
\label{SoftRotated}
\end{eqnarray}
The first three terms are MSSM-like terms, equivalent to the first
three terms in eq.~(\ref{SoftUnrot}). In fact, in analogy with the
MSSM, the coefficients of $|H'_1|^2$ and $|\widetilde L'_3|^2$ could
be defined in the rotated basis as the soft masses $m'^2_{H_1}$ and
$M'^2_{L_3}$ respectively, and the coefficient of $H'_1 H_2$ would be
the new bilinear soft term $B'\mu'$.  Notice however that the last two
terms violate R--Parity and tau lepton number, and correspond to the
last term in eq.~(\ref{SoftUnrot}). They are linear in the slepton and
therefore induce a non-zero tau sneutrino vev in the rotated basis
$\vev{\tilde\nu'_{\tau}}=v'_3/\sqrt{2}$.

Vacuum expectation values are calculated by minimizing the scalar
potential. In the original basis the extremization (or tadpole)
equations are
\begin{eqnarray}
t_1&=&(m_{H_1}^2+\mu^2)v_1-B\mu v_2-\mu\epsilon_3v_3+
\frac{1}{8}(g^2+g'^2)v_1(v_1^2-v_2^2+v_3^2)=0\,,
\nonumber \\
t_2&=&(m_{H_2}^2+\mu^2+\epsilon_3^2)v_2-B\mu v_1+B_2\epsilon_3v_3-
\frac{1}{8}(g^2+g'^2)v_2(v_1^2-v_2^2+v_3^2)=0\,,
\nonumber \\
t_3&=&(M_{L_3}^2+\epsilon_3^2)v_3-\mu\epsilon_3v_1+B_2\epsilon_3v_2+
\frac{1}{8}(g^2+g'^2)v_3(v_1^2-v_2^2+v_3^2)=0\,.
\label{eq:tadpoles}
\end{eqnarray}
The $t_i$ are the tree level tadpoles and they are equal to zero at
the minimum. The linear terms of the scalar potential are then
$V_{linear}=t_1\chi^0_1+t_2\chi^0_2+t_3\tilde\nu^R_{\tau}$, where
$\chi^0_i=\sqrt{2}Re(H^i_i)-v_i$ and
$\tilde\nu^R_{\tau}=\sqrt{2}Re(\tilde\nu_{\tau})-v_3$.  The first two
equations reduce to the MSSM minimization conditions after taking the
MSSM limit $\epsilon_3=v_3=0$, and in this case, the third equation is
satisfied trivially. Note that $\epsilon_3=0$ implies two solutions
for $v_3$ from the third tadpole in eq.~(\ref{eq:tadpoles}), from
which only $v_3=0$ is viable, since the second solution would imply
the existence of a massless isodoublet pseudoscalar majoron.

The first two tadpole equations in the rotated basis are
\begin{eqnarray}
t'_1&=&\mu'^2v'_1+{{m_{H_1}^2\mu^2+M_{L_3}^2\epsilon_3^2}\over{\mu'^2}}
v'_1-{{B\mu^2+B_2\epsilon_3^2}\over{\mu'}}v_2+
(m_{H_1}^2-M_{L_3}^2){{\epsilon_3\mu}\over{\mu'^2}}v'_3
\nonumber\\&&
+{\textstyle{1\over8}}(g^2+g'^2)v'_1(v'^2_1-v_2^2+v'^2_3)=0
\label{UnoTadpoleRot}\\ \nonumber\\
t'_2&=&\mu'^2v_2+m_{H_2}^2v_2-{{B\mu^2+B_2\epsilon_3^2}\over{\mu'}}v'_1
+(B_2-B){{\epsilon_3\mu}\over{\mu'}}v'_3
\nonumber\\&&
-{\textstyle{1\over8}}(g^2+g'^2)v_2(v'^2_1-v_2^2+v'^2_3)=0
\label{TwoTadpoleRot}
\end{eqnarray}
where $\vev{H'_1}=v'_1/\sqrt{2}$ and $\vev{ L'_3} =v'_3/\sqrt{2}$ with
$v'_1=(\mu v_1-\epsilon_3v_3)/\mu'$ and $v'_3=(\epsilon_3v_1+\mu
v_3)/\mu'$, as suggested by eq.~(\ref{eq:rotation}). These two
equations resemble the MSSM minimization conditions when we set
$v'_3=0$. The third tadpole equation is
\begin{eqnarray}
t'_3&=&(m_{H_1}^2-M_{L_3}^2){{\epsilon_3\mu}\over{\mu'^2}}v'_1
+(B_2-B){{\epsilon_3\mu}\over{\mu'}}v_2
+{{m_{H_1}^2\epsilon_3^2+M_{L_3}^2\mu^2}\over{\mu'^2}}v'_3
\nonumber\\&&
+{\textstyle{1\over8}}(g^2+g'^2)v'_3(v'^2_1-v_2^2+v'^2_3)=0
\label{tadpoleiii}
\end{eqnarray}
In this equation we note that $v'_3=0$ if $\Delta m^2 \equiv
m_{H_1}^2-M_{L_3}^2=0$ and $\Delta B \equiv B_2-B=0$ at the weak
scale. In supergravity models with universality of scalar soft masses
and bilinear mass parameters we have $\Delta m^2=0$ and $\Delta B=0$
at the unification scale $M_{GUT} \approx 2 \times 10^{16}$ GeV, but
radiative corrections lifts this degeneracy due to the running of the
renormalization group equations (RGE) between $M_{GUT}$ and
$M_{weak}$. In the approximation where $\Delta m^2$ and $\Delta B$ are
small we find that $v'_3$ is also small and in first approximation
given by
\begin{equation}
v'_3\approx -{{\epsilon_3\mu}\over{\mu'^2m_{\tilde\nu^0_{\tau}}^2}}
\left(v'_1\Delta m^2+\mu'v_2\Delta B\right)
\label{App_v3p}
\end{equation}
where we have introduced
\begin{equation}
m_{\tilde\nu^0_{\tau}}^2\equiv {{m_{H_1}^2\epsilon_3^2+M_{L_3}^2\mu^2}
\over{\mu'^2}}+{\textstyle{1\over8}}(g^2+g'^2)(v'^2_1-v_2^2)
\label{sneumassMSSM}
\end{equation}
which reduces to the tau sneutrino mass in the MSSM when we set
$\epsilon_3=0$.  Note that \eq{App_v3p} implies that the
R--parity-violating effects induced by $v'_3$ are {\sl calculable} in
terms of the primordial effective R--parity-violating $\epsilon$.

\section{Neutrino Mass}

The presence of tau lepton number and BRpV terms, characterized by the
parameters $\epsilon_3$ and $v_3$, leads to a mixing between
neutralinos and the tau neutrino \cite{rossarca}, as a result of which
the tau neutrino acquires a mass $m_{\nu_{\tau}}$. In the original
basis, where $(\psi^0)^T=
(-i\lambda',-i\lambda^3,\widetilde{H}_1^1,\widetilde{H}_2^2,\nu_{\tau})$,
the scalar potential contains the following mass terms
\begin{equation} 
{\cal L}_m=-1/2 (\psi^0)^T{\bf M}_N\psi^0+h.c.   
\label{eq:NeuMLag} 
\end{equation} 
where the neutralino/neutrino mass matrix is 
\begin{equation} 
{\bf M}_N=\left[  
\begin{array}{ccccc}  
M^{\prime } & 0 & -\frac 12 g^{\prime }v_1 & \frac 12g^{\prime }v_2 & -\frac  
12g^{\prime }v_3 \\   
0 & M & \frac 12gv_1 & -\frac 12gv_2 & \frac 12gv_3 \\   
-\frac 12g^{\prime }v_1 & \frac 12gv_1 & 0 & -\mu  &x 0 \\   
\frac 12g^{\prime }v_2 & -\frac 12gv_2 & -\mu  & 0 & \epsilon _3 \\   
-\frac 12g^{\prime }v_3 & \frac 12gv_3 & 0 & \epsilon _3 & 0  
\end{array}  
\right] 
\label{eq:NeuM5x5} 
\end{equation} 
where $M$ and $M'$ are the $SU(2)$ and $U(1)$ gaugino masses. It can
be seen from eq.~(\ref{eq:NeuM5x5}) that mixings between tau neutrino
and neutralinos are proportional to $\epsilon_3$ and $v_3$. Naively
one could think that, due to the strong experimental constraint on the
tau neutrino mass, the parameters $\epsilon_3$ and $v_3$ should be
very small with respect to the weak scale and, in fact, this has often
been claimed as a way to dismiss the phenomenological relevance of
R--parity violation. However, the cosmological critical density bound
\mnt $\lsim 92 \Omega h^2$ eV only holds if neutrinos are stable.  In
the present BRpV model (where there is no majoron) the \nt can decay
into 3 neutrinos, via the neutral current \cite{2227}, or by slepton
exchanges. This mechanism may be efficient in reducing the relic \nt
abundance below the required level, as long as \nt is heavier than
about 100 keV or so. On the other hand primordial Big-Bang
nucleosynthesis implies that \nt is lighter than about an Mev or
so. Thus, in addition to the electron-volt neutrino mass range, we
obtain another region, say between .1 to 1 MeV where heavy \nt masses
are cosmologically consistent in the BRpV model. Needless to say, in
the spontaneous breaking version of the model all masses up to the LEP
limit are cosmologically consistent due to the majoron-induced decay
and annihilation channels \cite{DRPV}.

\begin{figure}
\centerline{\protect\hbox{\psfig{file=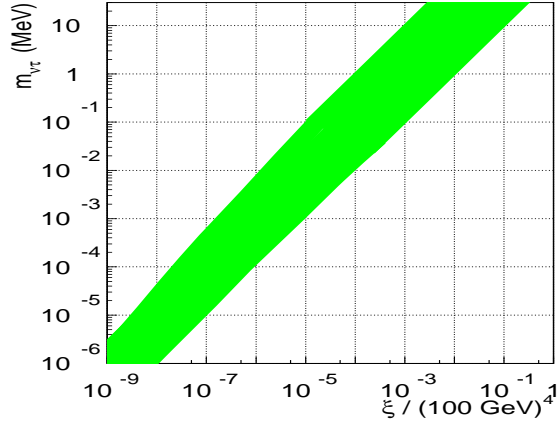,height=6cm,width=0.7\textwidth}}}
\caption{Tau neutrino mass as a function of the effective
R--parity-violating parameter $\xi\equiv(\epsilon_3v_1+\mu v_3)^2$.}
\label{mnutau_ev}
\end{figure}

Let us now compare the cosmologically allowed values of the tau
neutrino mass with the theoretically predicted ones. In order to do
this we embed our MSSM--BRpV model into supergravity, with
universality of scalar ($m_0$), gaugino ($M_{1/2}$), bilinear ($B$),
and trilinear ($A$) soft mass parameters at the unification scale $M_X
\approx 2 \times 10^{16}$ GeV \cite{epsrad}. The expected \mnt values
are illustrated in Fig.~\ref{mnutau_ev}, where we have imposed the
radiative breaking of the electroweak symmetry by minimizing the
scalar potential with the aid of one--loop tadpole equations. We have
made a scan over the parameter space, including the BRpV parameters
$\epsilon_3$ and $v_3$, imposing the LEP limit on \mnt and that the
supersymmetric particles are not too light.
As one can see there is a strict correlation between the neutrino mass
and the magnitude of R--parity-violation given by $\epsilon_3$ and
$v_3$ which need not be small. We have explicitly verified that
$|\epsilon_3|$ can be as large as 400 GeV and that $|v_3|$ can be
close to 100 GeV without conflicting with laboratory limits on the tau
neutrino mass. An obvious question arises at this stage: how can we
get a small neutrino mass, if so desired?  The answer lies in the fact
that the induced neutrino mass is $\propto(\epsilon_3v_1+\mu v_3)^2$,
and this last combination is what needs to be small. One can see that
the contributions to \mnt coming from Higgsino and gaugino mixing,
which are proportional to $\epsilon_3$ and $v_3$ may nearly cancel,
leading to a mass that can be very small, in the eV or sub-eV
range. How natural is such a cancellation? We have found that if our
model is unified a la supergravity with universality of scalar and
bilinear soft mass parameters, the combination $(\epsilon_3v_1+\mu
v_3)$ is radiatively induced, and therefore, naturally small.

In order to appreciate this better we rewrite neutral mass matrix in
eq.~(\ref{eq:NeuM5x5}) in the rotated basis. This corresponds to the
substitution $(v_1,v_3,\epsilon_3,\mu) \to (v'_1,v'_3,0,\mu')$. In
this basis the $\epsilon$ term is not present, and the only source of
mixing responsible for the neutrino mass is the vev $v'_3$. In first
approximation, valid when $v'_3$ is small, we get
\begin{equation}
m_{\nu_{\tau}} \approx-{{(g^2M+g'^2M')\mu'^2v'^2_3}\over{
4MM'\mu'^2-2(g^2M+g'^2M')v'_1v_2\mu'}}
\label{mNeutrinoApp}
\end{equation}
Now solving the renormalization group equations for the soft mass
parameters $m^2_{H_1}$, $m^2_{L_3}$, $B$, and $B_2$, in first
approximation we get
\begin{eqnarray}
m_{H_1}^2-M_{L_3}^2&\approx&-{{3h_b^2}\over{8\pi^2}}
\left(m_{H_1}^2+M_Q^2+M_D^2+A_D^2\right)\ln{{M_{GUT}}\over{m_Z}}
\nonumber\\
B_2-B&\approx&{{3h_b^2}\over{8\pi^2}}A_D\ln{{M_{GUT}}\over{m_Z}}
\label{m2BDiff}
\end{eqnarray}
Using eq.~(\ref{App_v3p}) we can show that the sneutrino vev $v'_3$
given through $\xi\equiv(\epsilon_3v_1+\mu v_3)^2=(\mu'v'_3)^2$ is
radiatively generated, with a maximum value of few GeV or so. The
resulting tau neutrino mass is given by
\begin{equation}
m_{\nu_{\tau}}\approx{{
\left[\mu'v_2A_D-v'_1\left(m_{H_1}^2+M_Q^2+M_D^2+A_D^2\right)\right]^2
}\over{\Big[2v'_1v_2-4MM'\mu'/(g^2M+g'^2M')\Big]
\mu'm_{\tilde\nu^0_{\tau}}^2}}
\left({{\epsilon_3\mu}\over{\mu'^2}}\right)^2
\left({{3h_b^2}\over{8\pi^2}}\ln{{M_{GUT}}\over{m_Z}}\right)^2
\label{mNeutriApp2}
\end{equation}
This mass can be further approximated by
\begin{equation}
m_{\nu_{\tau}}\approx{{m_Z^2}\over{M_{SUSY}}}
\left({{\epsilon_3}\over{M_{SUSY}}}\right)^2h_b^4 \sim 1\,{\mathrm{eV}}
\label{mNeutriApp3}
\end{equation}
where we have explicitly indicated that 1 eV is a perfectly viable
\mnt value in this model. This was obtained for $M_{SUSY} \sim
\epsilon_3 \sim m_Z$ and $h_b\sim 10^{-2}$ \cite{epsrad}. Therefore,
\mnt can be naturally small, even though the R--parity-violating
parameters are large. The actual scale on neutrino mass can, of
course, be larger, as the smallness of \mnt is tightly related to our
soft SUSY breaking terms universality assumption at the unification
scale.

\section{Unification of Couplings}

Unification of the gauge couplings in our model works basically as in
the MSSM. In contrast, Yukawa coupling unification is rather
different. To carry out this discussion we start from the basic
superpotential in \eq{eq:Wsuppot}.  Similarly to neutralino-neutrino
mixing, charginos also mix with the tau lepton, forming a set of three
charged fermions $F_i^{\pm}$, $i=1,2,3$. In the original basis where
$\psi^{+T}=(-i\lambda^+,\widetilde H_2^1,\tau_R^+)$ and
$\psi^{-T}=(-i\lambda^-,\widetilde H_1^2,\tau_L^-)$, the charged
fermion mass terms in the Lagrangian are ${\cal L}_m=-\psi^{-T}{\bf
M_C}\psi^+$, with the mass matrix given by
\begin{equation} 
{\bf M_C}=\left[\matrix{ 
M & {\textstyle{1\over{\sqrt{2}}}}gv_2 & 0 \cr 
{\textstyle{1\over{\sqrt{2}}}}gv_1 & \mu &  
-{\textstyle{1\over{\sqrt{2}}}}h_{\tau}v_3 \cr 
{\textstyle{1\over{\sqrt{2}}}}gv_3 & -\epsilon_3 & 
{\textstyle{1\over{\sqrt{2}}}}h_{\tau}v_1}\right] 
\label{eq:ChaM6x6} 
\end{equation} 
As a result, the tau Yukawa coupling is not related to the tau mass by
the usual MSSM relation.  In contrast, $h_{\tau}$ depends now on the
parameters of the chargino sector $M$, $\mu$, and $\tan\beta$, as well
as the BRpV parameters $\epsilon_3$ and $v_3$, through a formula given
in ref.~\cite{ChaStau}. In addition, the top and bottom quark Yukawa
couplings are related to the quark masses in a way different from that
of the MSSM due to the non-zero value of $ v_3/v$, i.e.
\begin{equation}
m_t = h_t {v\over{\sqrt2}} \sin \beta \sin \theta\,,\qquad
m_b = h_b {v\over{\sqrt2}} \cos \beta \sin \theta 
\end{equation}
where $v=246$ GeV and we have defined $\cos\theta\equiv v_3/v$.

\begin{figure}
\centerline{\protect\hbox{ 
\psfig{file=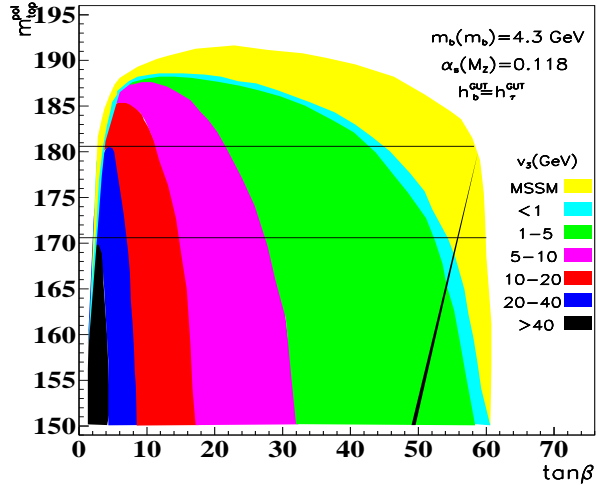,height=7.5truecm,width=9truecm}}}
\caption{Pole top quark mass as a function of $\tan\beta$ for
different values of the R--Parity violating parameter $|v_3|$. Bottom
quark and tau lepton Yukawa couplings are unified at $M_{GUT}$.}
\label{aretop}
\vglue -.5cm
\end{figure}

The re-scaling in the bottom quark Yukawa term ensures that the same
quark mass is obtained with the same Yukawa coupling in the two
basis. This re-scaling with respect to the MSSM is non-trivial and has
profound consequences in Yukawa unification, as shown in
Fig.~\ref{aretop}, taken from ref. \cite{YukUnif}.
In this figure we observe that bottom--tau Yukawa unification can be
achieved at any value of $\tan\beta$ by choosing appropriately the
value of $v_3$ \cite{YukUnif}. The horizontal lines correspond to the
$1\sigma$ experimental determination of $m_t$.  The plot in
Fig.~\ref{aretop} is obtained through a scan over parameter space such
that points which satisfy $h_b(M_{GUT})=h_{\tau}(M_{GUT})$ within
$1\%$ are kept, where $M_{GUT}$ is the gauge coupling unification
scale.  Each selected point is placed in one of the regions of
Fig.~\ref{aretop} according to its $|v_3|$ value.  However, one can
see that, for very large $v_3 \gsim 40$ GeV the large $\tan\beta$
solution is ruled out.  Points with top-bottom-tau unification are
concentrated in the diagonal line at high values of $\tan\beta$,
analogously to the MSSM case \cite{YukUnif}.

In summary, BRpV is the simplest way to introduce R--Parity violation
to the MSSM. The model can be embedded into Supergravity models with
universality of scalar, gaugino, bilinear and trilinear soft mass
parameters. In this case, the induced tau neutrino mass arises
radiatively and is naturally small. The BRpV parameters $\epsilon_3$
and $v_3$ need not be small and can be easily of the order of $m_Z$.
Another important feature is that BRpV changes the relation between
the Yukawa couplings and the masses of the top and bottom quarks and
the tau lepton. As a result, bottom-tau Yukawa unification can be
achieved for any $\tan\beta$ value, provided we choose appropriately
the value of the sneutrino vev $v_3$.  Even in the unlikely limit
where the tau neutrino is massless with $\epsilon_3 \neq 0$ (which
corresponds to having universality of soft mass parameters {\sl at the
weak scale!}, which is not natural) R--Parity is not conserved. In
fact, even though the neutralinos decouple from the tau neutrino, the
lightest neutralino decays for example to $b\overline b \nu_{\tau}$
through an intermediate sbottom due to the last term in
\eq{WsuppotP}. Thus R--parity violation can be sizeable even if
neutrinos turn out to be very light, as indicated by present solar and
atmospheric neutrino data. Some of the phenomenological implications
of the model have been discussed in ref. \cite{BRpVtalkphen}.

\section*{Acknowledgments:} 

This work was supported by DGICYT grant PB95-1077 and by the EEC under
the TMR contract ERBFMRX-CT96-0090.

\end{document}